\documentclass[prl,floatfix,twocolumn,final]{revtex4}
\usepackage{graphics}
\usepackage{graphicx}

\newcommand{\eqref}[1]{(\ref{#1})}
\renewcommand{\d}{{\, \mathrm{d}}}

\newcommand{\exv}[1]{\left\langle{\: #1 \:}\right\rangle}
\newcommand{\exvs}[1]{\langle{#1}\rangle}

\newcommand{\be}{\begin{equation}}
\newcommand{\ee}[1]{\label{#1} \end{equation}}
\newcommand{\ba}{\begin{eqnarray}}
\newcommand{\ea}[1]{\label{#1} \end{eqnarray}}
\newcommand{\nl}{\nonumber \\}

\begin{document}

\title{ {\bf Mass gap from pressure inequalities}}

\author{Tam\'as S. Bir\'o
}
\affiliation{
 KFKI Research Institute for Particle and Nuclear Physics Budapest
}
\author{Andr\'as L\'aszl\'o
}
\affiliation{
 KFKI Research Institute for Particle and Nuclear Physics Budapest
}
\author{P\'eter V\'an
}
\affiliation{
 KFKI Research Institute for Particle and Nuclear Physics Budapest,
 BCCS Bergen Computational Physics Laboratory, Bergen
}

\date{\today}

\begin{abstract}
We prove that a temperature independent mass distribution
is identically zero below a mass threshold (mass gap) value, if the
pressure satisfies certain inequalities. This supports the finding
of a minimal mass in quark matter equation of state by numerical 
estimates and by substitution of analytic formulas.
We present a few inequalities for the mass distribution based on the Markov 
inequality.
\end{abstract}

\maketitle

%%%%%%%%%%%%%%%%%%%%%%%%% SECTION 1 %%%%%%%%%%%%%%%%%%%%%%%%%%
\section{Introduction}

Quark matter, searched for in relativistic heavy ion collisions, reveals
itself in signatures on observed hadron spectra which are (can best be)
interpreted in terms of quark level properties. In particular scaling
of the elliptic flow component $v_2$ with the constituent quark content
of the finally observed mesons and baryons \cite{v2} and successful
description of $p_T$-spectra of pions and antiprotons using quark
coalescence rules for hadron building \cite{ourJPG} utilize the fast
hadronization concept of quark redistribution.
Albeit this simple idea brings also problems with it, e.g. in dealing
with energy conservation and entropy non-decreasing, these issues can be
resolved by using a distributed mass quasiparticle model for quark
matter \cite{ourPRChep}, and are in accord with the quark matter
equation of state obtained in lattice QCD calculations \cite{ourPLB}. 
The surmised mass distribution gives
rise to specific equation of state (pressure as a function of temperature, $p(T)$),
and reversed, a mass distribution may be outlined from knowledge on the
$p(T)$ curve.

While traditional, fixed mass quasiparticle
models already succeed to describe the equation of state obtained in
lattice QCD \cite{Quasiptl}, those mass values are themselves
temperature dependent. Furthermore a temperature dependent width
is associated to the quasiparticle mass, too \cite{Peshier}. This is unavoidable,
as it can be easily seen  from the following comparison of pressures
obtained from a mass distribution and from perturbative QCD in the high
temperature limit:
\ba
 \frac{p(T)}{p_{{\rm id}}(0,T)} & = & 
 1 - \frac{\exvs{m^2}}{4T^2}+ \left(\frac{3}{4}-\gamma\right)\frac{\exvs{m^4}}{16T^4}
 \nl
 & & + \, \frac{\exvs{m^4\ln(2T/m)}}{16T^4} +\ldots, \nl
 \nl
 \frac{p_{{\rm pQCD}}}{p_{{\rm id}}(0,T)} &=& 1 - a_2g^2 +a_4g^4 +b_4g^4\ln\frac{2\pi T}{\Lambda}+\ldots 
\ea{pQCD}
where the ideal gas pressure for relativistic particles with mass $m$ defines
the scaling function
\be
 \Phi\left(\frac{m}{T}\right) = \frac{p_{{\rm id}}(m,T)}{p_{{\rm id}}(0,T)} =  \frac{1}{2}\; 
 \left(\frac{m}{T} \right)^2 K_2\left(\frac{m}{T} \right).
\ee{RELID}
We note that $\Phi(x)$ can also be obtained for Bose or Fermi
distributions instead of the Boltzmann one; the numerical difference
is overall minor, less than six per cent. 
The basic assumption, $\exv{m^2}\sim g^2T^2$
sets the scale for a simplified treatment of the quark matter pressure at high
temperature.
The comparison reveals that $\exv{m^4} \ne \exv{m^2}^2$, whence the necessity of
a width in the mass distribution emerges. The temperature dependence of the
pressure ratio to the massless ideal gas value is concentrated on the
temperature dependence of the coupling constant: $g=g(T)$ in the traditional
interpretation. We have recently pursued an alternative approach to the
quasiparticle mass distribution in quark matter \cite{ourJPG,ourPRChep}, where
a temperature independent $w(m)$ distribution is reconstructed from the
pressure ratio $\sigma(T)=p(T)/p_{{\rm id}}(0,T)$ curve:
\be
 \sigma(T) = \int_0^{\infty} w(m) \: \Phi\left(\frac{m}{T}\right) \, \d m.
\ee{PRESS}
This is possible only with a single $w(m)$ curve for each exactly known $\sigma(T)$.
Such $w(m)$ distributions show diverging expectation values for positive
powers of the mass (like $\exv{m^2}$) signaling a high mass tail
not falling faster than $\sim m^{-3}$. Another remarkable property of this approach
is that it indicates a temperature independent threshold (smallest mass) in the
$w(m)$ spectrum for lattice QCD pressure data\cite{FODOR,BIELEFELD}.

The pressure is, however, not known analytically, the numerical results are
smeared with error bars. This problem is more severe in the light of the
fact that eq.(\ref{PRESS}) is an integral transformation (related to the
Meijer K-transformation, a generalization of the Laplace transformation).
There is no mathematical guarantee that by inverting such a transformation one
obtains close results for $w(m)$ from close functions for $\sigma(T)$.
In fact this is known as the ''inverse imaging problem'' \cite{IMAG}.

\begin{figure}[htb]

\includegraphics[width=0.28\textwidth,angle=-90]{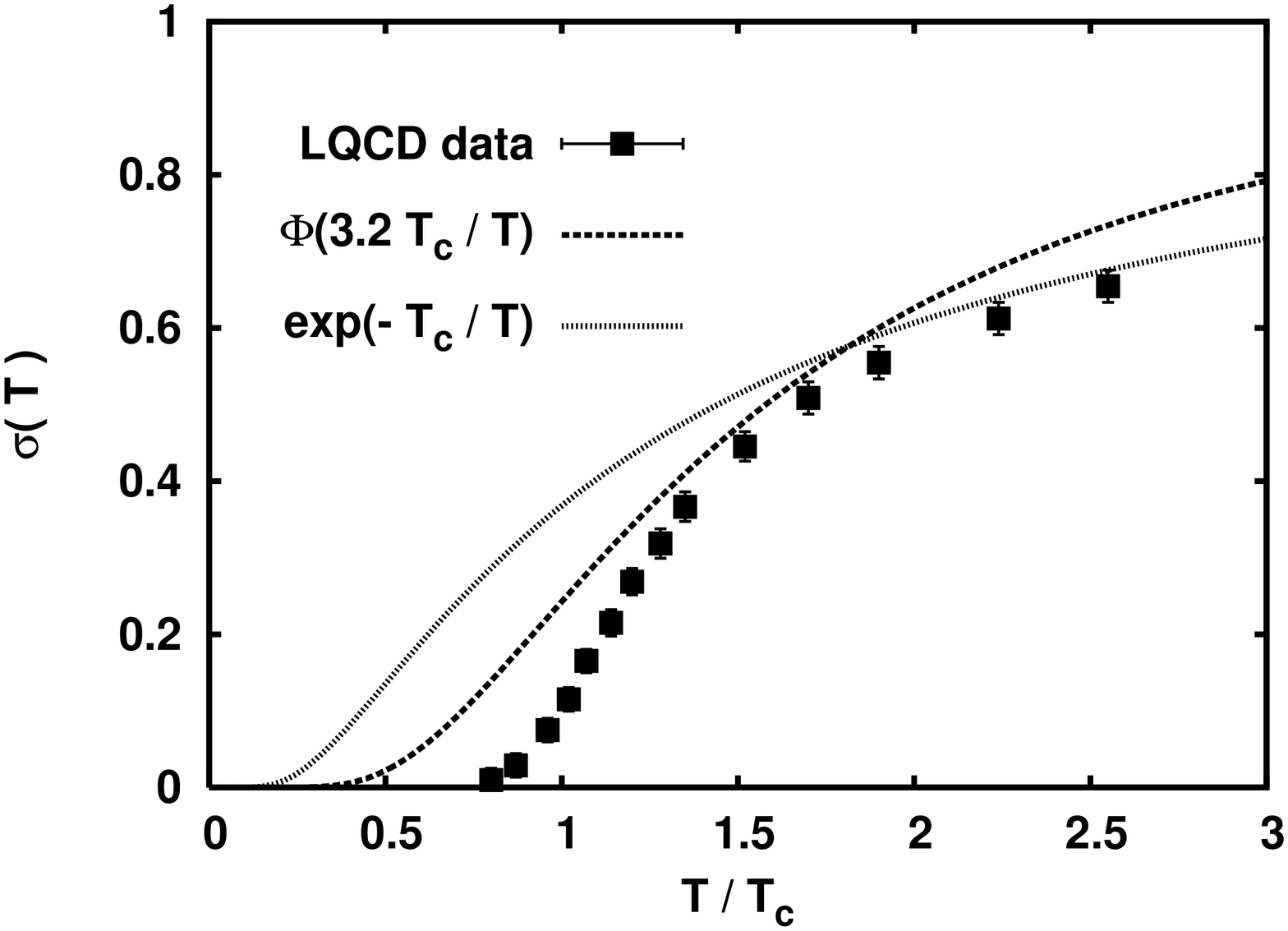}
\includegraphics[width=0.28\textwidth,angle=-90]{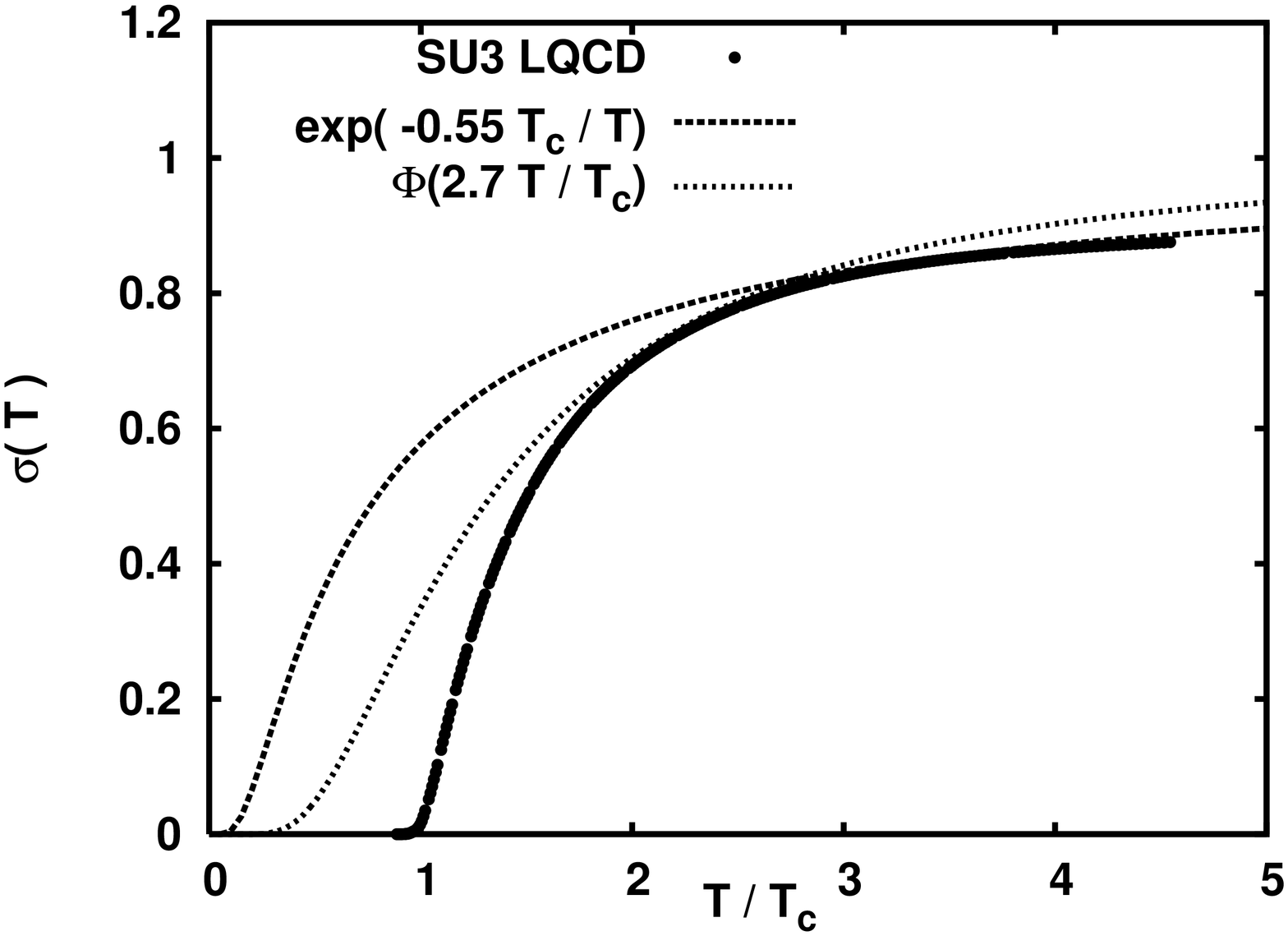}

\caption{ \label{FIG0}
  Pressure curve estimators for the data from lattice QCD simulation of Ref.\cite{FODOR}
  (2+1 flavor QCD) and of \cite{SU3} (pure SU(3) gauge theory).
}
\end{figure}

Our goal in the present paper is to support knowledge about a $T$-independent $w(m)$
mass distribution when the pressure $p(T)$ satisfies certain inequalities.
In particular we prove that if the pressure $p(T)$ is below the corresponding
ideal gas pressure with a given mass $M_0$ at {\em all temperatures}, then the
mass distribution is exactly zero for all masses below $M_0$. 
For inequalities with other than ideal gas pressure curves as
estimators we apply the Markov inequality for probability measures,
which directly offers upper bounds on the integrated probability
$P(M)=\int_0^M w(m) dm$.  
It turns out that the appearance and value of the highest possible
$M$ for which $P(M)=0$, the mass gap value $M_0$, 
is connected to the low temperature behavior of $\sigma(T)$.
Two particular estimators for $\sigma(T)$, namely $\Phi(M_0/T)$ with $M_0=3.2 T_c$ and
$\exp(-T_c/T)$ are compared to 2+1 flavor lattice QCD scaled pressure data
in Figure \ref{FIG0} (top) and to pure SU(3) lattice gauge theory data (bottom) with
$M_0=2.7T_c$ and $\lambda=0.55T_c$.
Of course the temperature scales are different, $T_c\approx 165$ MeV in the first,
$T_c \approx 260$ MeV in the second case. 
These examples are important for gaining a physical insight
into the Markov inequality discussed below.

%%%%%%%%%%%%%%%%%%%%%%%%%%%%%%% FIG %%%%%%%%%%%%%%%%%%%%%%%%%%%%%%
%%%%%%%%%%%%%%%%%%%%%%%%%%%%%%%%%%%%%%%%%%%%%%%%%%%%%%%%%%%%%%%%%%

%%%%%%%%%%%%%%%%%%%%%%%%%%%%%%% FIG %%%%%%%%%%%%%%%%%%%%%%%%%%%%%%
%\begin{figure}[htb]

%\includegraphics[width=0.45\textwidth,angle=-90]{fig1.eps}
%\includegraphics[width=0.3\textwidth,angle=-90]{fig1.eps}

%\caption{ \label{FIG1}
%  The Boltzmann-kernel, $\Phi(x)$, for obtaining the pressure from a continuous mass distribution.
%  Defined for positive real arguments it is continuous, strict monotonic, decreasing from the
%  value $\Phi(0)=1$ towards zero.
%}
%\end{figure}
%%%%%%%%%%%%%%%%%%%%%%%%%%%%%%%%%%%%%%%%%%%%%%%%%%%%%%%%%%%%%%%%%%

%%%%%%%%%%%%%%%%% moments %%%%%%%%%%%%%%%%
\section{Markov inequality and mass gap}

In this section we derive the mass gap based on the Markov inequality
and use this approach to estimate upper bounds on the integrated probability
for the mass being lower than a given value.
The general form of the Markov inequality is given by \cite{MARKOV}
\be
 \mu\left(\left\{x\in X: f(x) \ge t \right\} \right) \le \frac{1}{g(t)} 
  \int\limits_{x\in X} g\left(\,f(x)\,\right) \d\mu(x)
\ee{MARKOV}
with measure $\mu$, a real valued $\mu$-measurable function $f$,
and a monotonic growing non-negative measurable real function $g$. 
The proof, based on the monotonity of integration, 
can be presented in a few lines. For a non-negative and monotonic growing
function $g(t) \le g(f(x))$ for $t \le f(x)$. We obtain
\be
 g(t)\!\!\!\!\!\!\int\limits_{f(x)\ge t}\!\!\!\!\!\!\d\mu(x)  \: =  
 \!\!\!\!\!\!\int\limits_{f(x)\ge t}\!\!\!\!g(t) \d\mu(x) \: \le 
 \!\!\!\!\!\!\int\limits_{f(x)\ge t}\!\!\!\!g\left(\, f(x)\, \right) \d\mu(x).
\ee{MPROOF1}
This quantity can be bounded by 
\be
 \int\limits_{f(x)\ge t} g\left(\, f(x)\, \right) \d\mu(x) \: \le \:
 \int\limits_{x\in X} g\left(\, f(x)\, \right) \d\mu(x).
\ee{MPROOF2}
A division by $g(t)\ge 0$ delivers the original statement in eq.(\ref{MARKOV}).

In order to apply this inequality to the mass spectrum we choose $f(m)=tM/m$.
In this case
\be
 \mu \left(\frac{tM}{m} \ge t \right) = \int_0^{M} \d\mu(m),
\ee{WHATmu}
and the Markov inequality reads as
\be
 P(M) := \int_0^{M} \d\mu(m) \le \frac{1}{g(t)} \int_0^{\infty}g\left(\frac{tM}{m}\right) \d\mu(m).
\ee{OTHER-MARKOV}
For a continuum mass spectrum $\d\mu(m)=w(m)\d m$ can be chosen with $w(m)$ being the
probability density function. The generalized Markov inequality stated above is
valid for general probability measures\footnote{a measure normalized to one} 
$\mu$ possibly including bound state contributions.

%%%%%%%% ------------------------- %%%%%%%%%%%%%%%
\subsection{Power law estimate}

In the following we discuss a few examples for monotonic rising functions $g(z)$, which
allow us to draw some conclusions about the integrated probability for masses below
$M$.  Applying the special form of $g(t)=t^n$ we arrive at
\be
 P(M) \le \frac{1}{t^n} \int_0^{\infty}\left(\frac{tM}{m}\right)^n w(m) \d m,
\ee{PSMALLER}
whence we obtain:
\be
 P(M) \le M^n \int_0^{\infty} m^{-n} w(m) \d m.
\ee{ourMARKOV}
It is easy to see that the negative integral moments of the mass on the right hand side of
the above inequality are connected to the negative integral moments of scaled pressure
$\sigma(T)=p(T)/p_{{\rm id}}(0,T)$. The final inequality for the probability of having
masses smaller than $M$ is given by
\be
 P(M) \le M^n \frac{\int_0^{\infty} T^{-n-1} \sigma(T) \d T}{\int_0^{\infty}x^{n-1}\Phi(x)\d x}.
\ee{AHA}

%%%%%%%%%%%%%%%%%%%%%%%%%%%%%%% FIG %%%%%%%%%%%%%%%%%%%%%%%%%%%%%%
\begin{figure}[thb]

\includegraphics[width=0.28\textwidth,angle=-90]{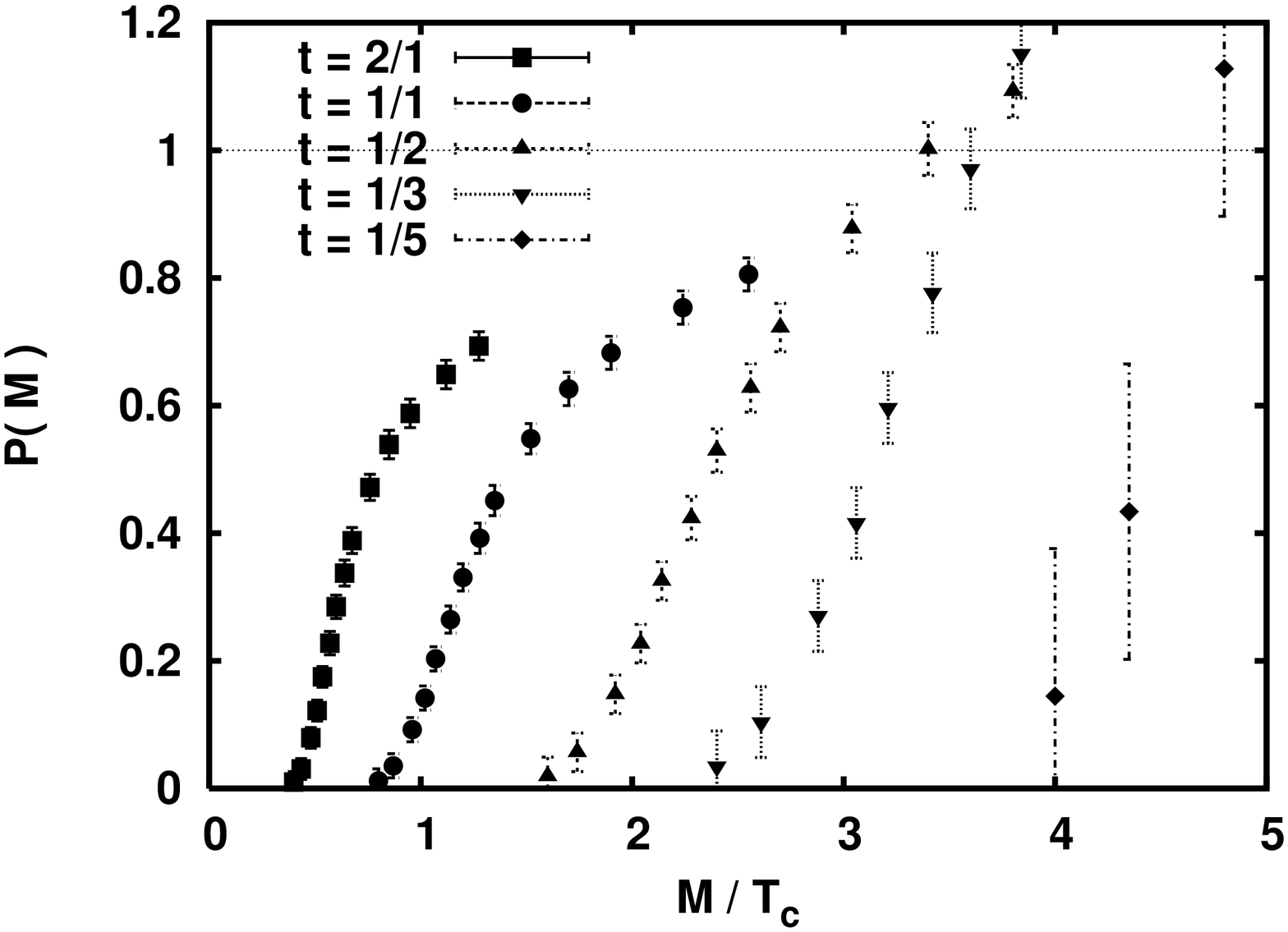}
\includegraphics[width=0.28\textwidth,angle=-90]{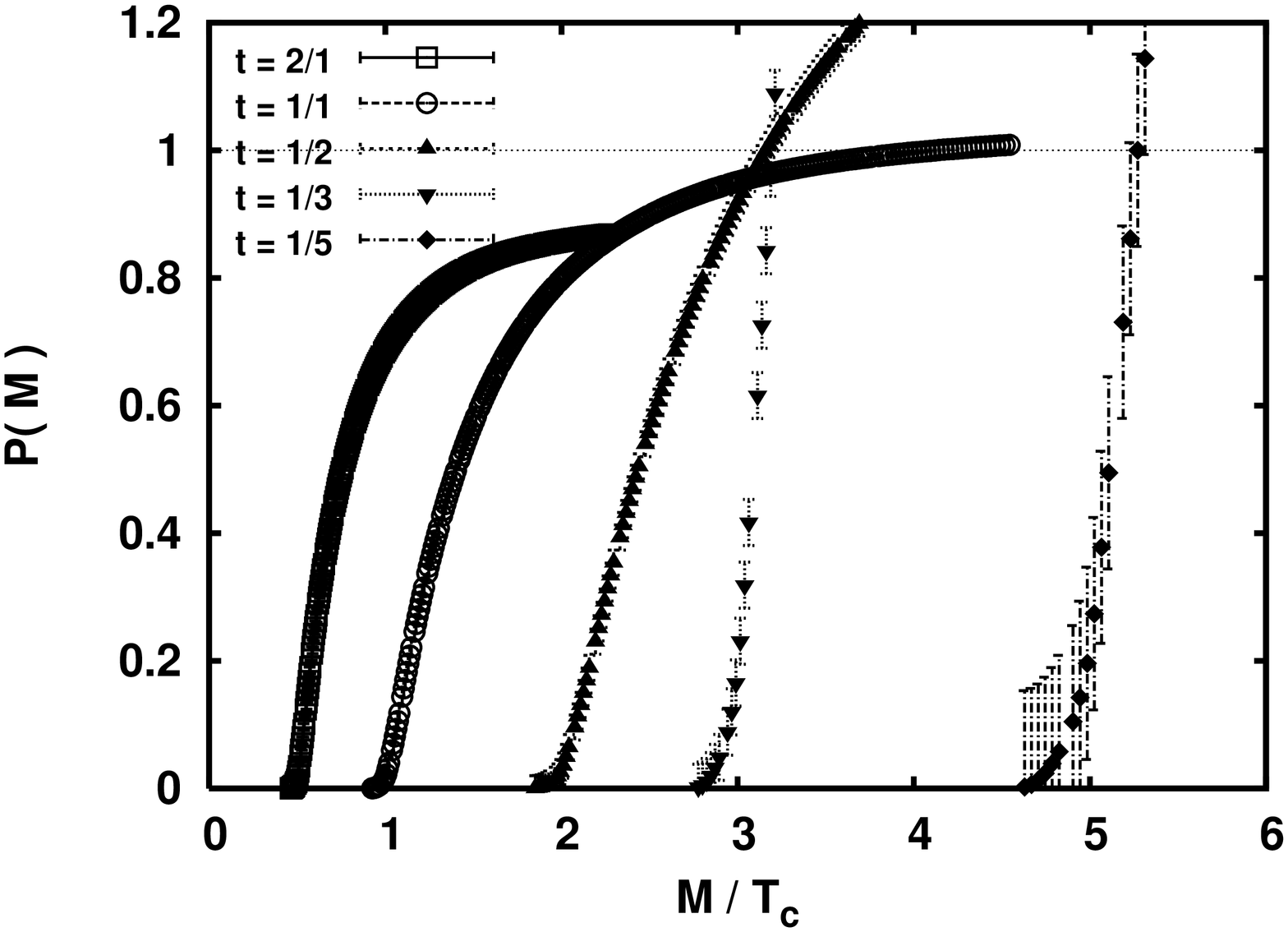}

\caption{ \label{FIG-PM}
  Upper bounds for the integrated probability $P(M)$ of masses lower than $M$,
  based on 2+1 flavor lattice QCD eos data\cite{FODOR} (top) and for pure
  SU(3) gauge theory data (bottom). Each data line belongs
  to a fixed value of the scaling parameter $t$ occurring in eq.(\ref{STRIKING}), see legend.
  In the SU(3) case a constant error of $0.02$ was assumed in the original
  $p/T^4$ data.
}
\end{figure}
%%%%%%%%%%%%%%%%%%%%%%%%%%%%%%%%%%%%%%%%%%%%%%%%%%%%%%%%%%%%%%%%%%
\subsubsection{Upper bound by a fixed mass ideal gas}

Let us apply this result to the simplest majorant, that of a fixed mass relativistic ideal gas.
In this case $\sigma(T)\le \Phi(M_0/T)$ with some $M_0$ (cf. dashed line in Figure \ref{FIG0}). 
Equation (\ref{AHA}) leads to
\be
 P(M)\: \le \: M^n
 \frac{\int_0^{\infty}T^{-n-1} \, \Phi\!\left(\frac{M_0}{T}\right) \d T }{\int_0^{\infty}x^{n-1}\Phi(x) \d x}
 \: \le \: \left(\frac{M}{M_0}\right)^n
\ee{IDEAL}
in this case. Should it hold for arbitrary high $n$, the right hand side of this inequality
is zero for all $M<M_0$ and divergent for $M>M_0$. In the second case it is not restrictive,
since $P(M)<1$ anyway, in the first case this means a mass gap up to $M_0$.
We note that this conclusion holds for a general non-negative $\Phi(x)$, for which the 
integrals 
in eq.(\ref{IDEAL}) are finite for all $n>0$. Thus the Bose-Einstein or the Fermi-Dirac
distribution could as well be applied instead of the Boltzmann one.

\subsubsection{Exponential upper bound}

Another possible majorant is the exponential function, $\sigma(T) \le \exp(-\lambda/T)$
(cf. the dotted line in Figure \ref{FIG0} for $\lambda=T_c$). 
In this case eq.(\ref{AHA}) delivers
\be
 P(M) \le \left(\frac{M}{2\lambda}\right)^n \frac{2\Gamma(n)}{\Gamma(2+n/2)\Gamma(n/2)}.
\ee{EXP}
The large $n$ limit of this result is given by
\be
 P(M) \le \left(\frac{M}{\lambda}\right)^n 2\sqrt{\frac{2}{\pi}} 
  \frac{1}{n^{3/2}}
\ee{LARGE-N-EXP}
to leading order in $1/n$. Again the right hand side approaches zero for $M\le \lambda$
and diverges for $M > \lambda$. This points out a mass gap stretching to 
(and including) $\lambda$ from zero.

%%%%%%%%%%%%%%%%%%%%%%%%%%%%%%% FIG %%%%%%%%%%%%%%%%%%%%%%%%%%%%%%
\begin{figure}[hb]

\includegraphics[width=0.28\textwidth,angle=-90]{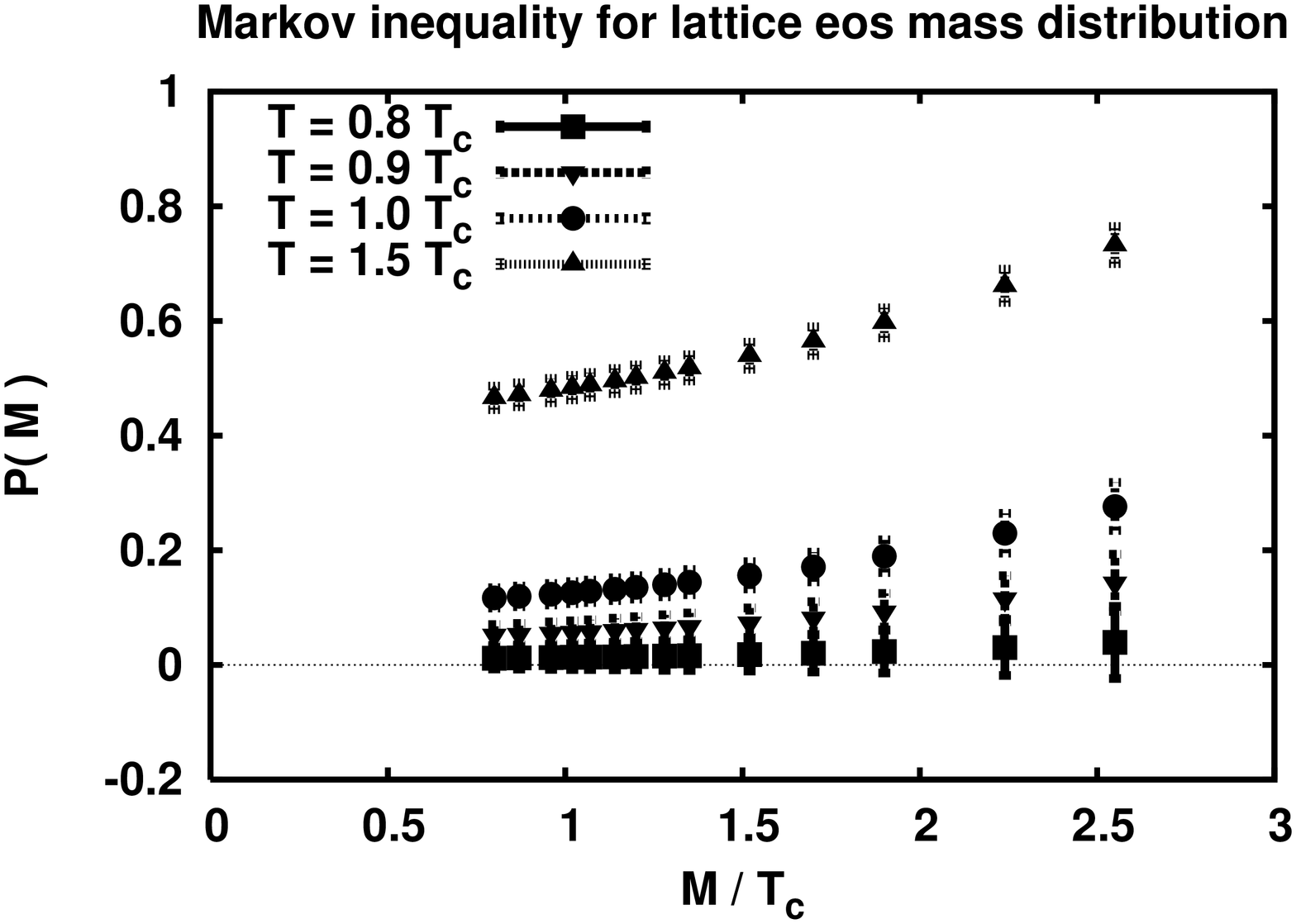}
\includegraphics[width=0.28\textwidth,angle=-90]{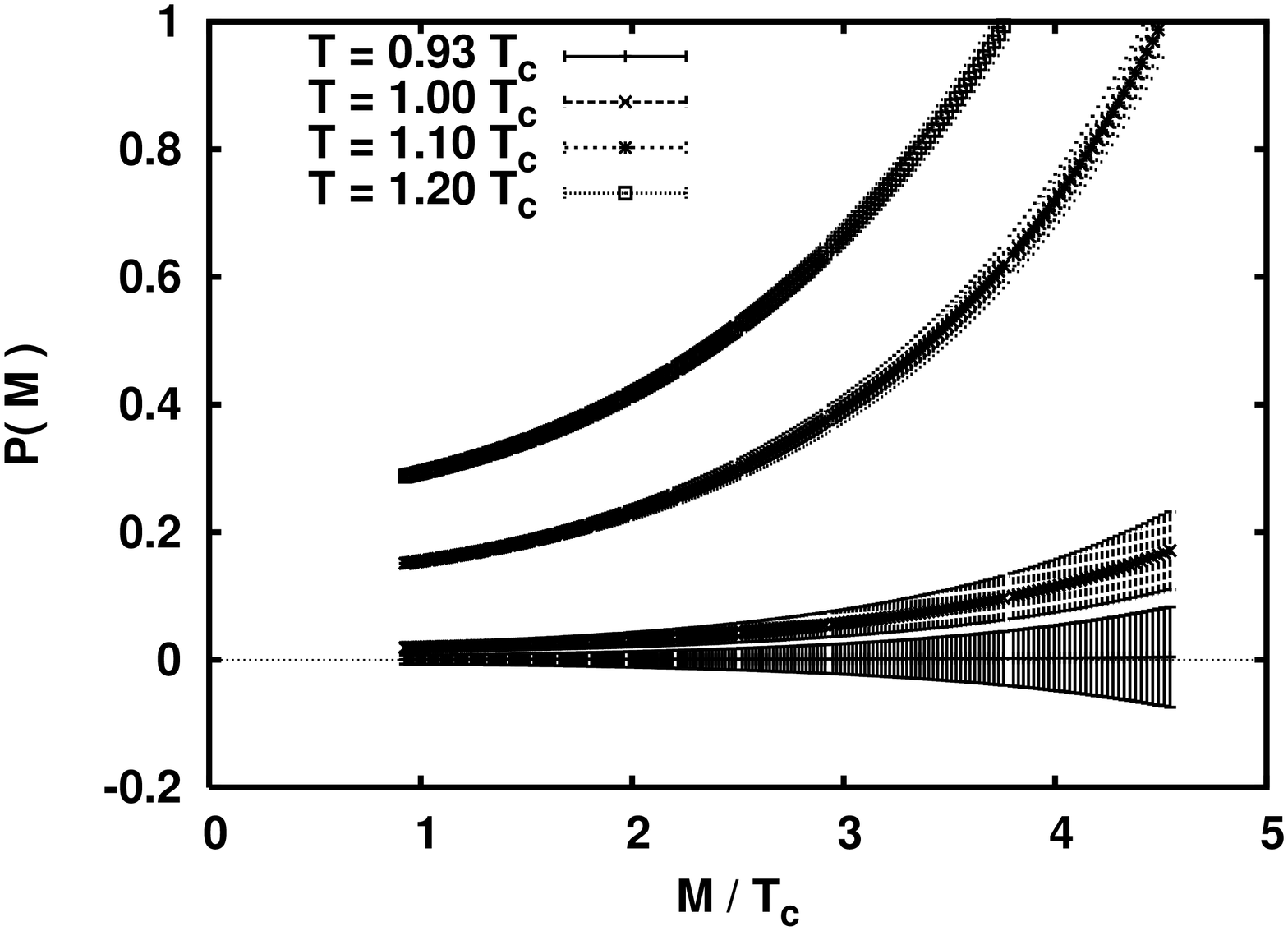}

\caption{ \label{M2}
  Upper bounds for the integrated probability of masses lower than $M$ based
  on eq.(\ref{PRACTICAL}) and
  on 2+1 flavor lattice QCD eos data\cite{FODOR}(top) and from Ref.[LPdata] (bottom)
  respectively.  
  The estimates belong to different temperatures  $T$ near $T_c$
  indicated in the legend. In the second case a constant error of $0.01$ was assumed
  in the original $p/T^4$ data.
}
\end{figure}
%%%%%%%%%%%%%%%%%%%%%%%%%%%%%%%%%%%%%%%%%%%%%%%%%%%%%%%%%%%%%%%%%%

\subsection{The pressure itself is a majorant}

The most striking inequality is obtained by using $g(t)=\Phi(1/t)$.
This function is also admissible, its rise from zero to one is strict monotonic.
Eq. (\ref{AHA}) leads to
\be
 P(M) \: \le \: \frac{\sigma(tM)}{\Phi(1/t)}.
\ee{STRIKING}
For $t=1$ using the numerical value $\Phi(1)\approx 0.81$ one arrives at
$P(M) \le 1.23 \sigma(M)$, which can be directly read off from numerical
simulation or theoretical predictions of $\sigma(T)$.
Figure \ref{FIG-PM} presents curves for different $t$-values (see legend), all being
an upper estimate for the integrated probability $P(M)$ in the respective cases of
2+1 flavor QCD and pure SU(3) gauge theory. The higher seems to be the starting
$M_0$ value for the rise of the upper bound on $P(M)$, the higher also
the magnification of the error bars. A secure estimate for the $P(M)\le 0.05$
is given for masses $M > 1.7 T_c = 280$ MeV for the 2+1 flavor QCD case,
while for $M > 7.2T_c = 1.9$ GeV for the pure SU(3) gauge case. While in the first case
this can be at best an avarage between quark and gluon-like quasiparticle masses,
in the second case should be close to observed glueball mass.
We note that using $\sigma(tM) \le \Phi(M_0/tM)$ in the $t \rightarrow 0$ limit
again a mass gap at $M_0$ follows from eq.(\ref{STRIKING}). In this respect
the use of different $g(z)$ functions in the Markov inequality does not 
matter\footnote{From practical viewpoint, however, in the $t\rightarrow 0$ limit
the error bars on the origional $p/T^4$ data are infinitely enlarged.}.

A related version of the inequality (\ref{STRIKING}) is obtained for $tM=T$,
$g(z)=\Phi(tM/zT)$.
The upper bound is obtained at any fixed $T$ as being
\be
 P(M) \: \le \: \frac{\sigma(T)}{\Phi(M/T)}
\ee{PRACTICAL}
Figure \ref{M2} plots upper bounds for $P(M)$ obtained using the
eq.(\ref{PRACTICAL}). The most restrictive are the lowest temperature
data for $\sigma(T)$, they are, however, also the most contaminated
by errors. It is probably safe to conclude that as much as
$90-95\%$ of the masses are above $1.5 T_c \approx 440$ MeV according to
these data.

Our mathematical treatment of the mass gap leaves the point $m=0$ in the
possible mass distribution as a special case. Assuming that there were
such a contribution of finite measure, i.e. $P(0)=a$ were a finite
value between zero and one, one concludes from the definition
eq.(\ref{PRESS}) that in this case $\sigma(T) \ge a$ would be. There is no
sign of such an indication in lattice QCD data.

Finally we note that there is a potential to use our method presented in this
letter in a context wider than quark matter: the quasiparticle test
based on the generalized Markov inequality can in principle be done for
any system with sufficiently known thermal equation of state. The estimate
for a lowest mass can then be checked against knowledge on the mass
spectrum obtained from the study of correlation functions.

%%%%%%%%%%%%%%%%%%%%%%%%%%%%% ACKNOWLEDGEMENT %%%%%%%%%%%%%%%%%

\vspace{7mm}
{\bf Acknowledgment}
This work has been supported by the Hungarian National Research Fund,
OTKA (T49466, T048483) and by the B\'olyai scholarship for P\'eter V\'an.
Discussions with Prof.~T.~Matolcsi, Prof.~T.~M\'ori and \'A.~Luk\'acs
are gratefully acknowledged.

%%%%%%%%%%%%%%%%%%%%%%%%%%%%% RFERENCES %%%%%%%%%%%%%%%%%%%%%%%

%%%%%%%%%%%%%%%%%%%%% BIBLIOGRAPHY %%%%%%%%%%%%%%%%%%%%%%%%%%%%%
%                 18.06.2006.
%
% ------------------------------------------------------

%%%%%%%%%%%%%%%%%%%%%%%%%%%%%% END %%%%%%%%%%%%%%%%%%%%%%%
\end{document}